\documentclass[nofootinbib,preprint,aps,superscriptaddress,showpacs,twocolumn]{revtex4-1}
\usepackage{graphicx}
\usepackage{caption}
\usepackage{tabularx}
\usepackage{amsmath}
\usepackage{color}
\usepackage{multirow}
\usepackage[FIGTOPCAP]{subfigure} %this packege with this option set the subfigure caption on the top of anf subfig.
\usepackage{array}
\newcolumntype{P}[1]{>{\centering\arraybackslash}p{#1}}
\newcolumntype{M}[1]{>{\centering\arraybackslash}m{#1}}

\newcommand{\be}{\begin{equation}}
\newcommand{\ee}{\end{equation}}

\newcolumntype{C}{>{\centering\arraybackslash}X}
\begin{document}

\title{{\bf Variational Calculations for Relativistic Two Dimensional Strongly Interacting Fermions: Application to 2D Liquid $^3He$ }}

\author{\bf H. Salehi}
\affiliation{\small{Department of Physics, Faculty of Science, Shahid Chamran University of Ahvaz, Ahvaz, Iran.}}
\author{\bf M. Mohammadi Sabet}
\affiliation{\small{Department of physics, Faculty of science, Ilam University, Ilam, Iran.}}
\author{\bf S. Mizani}
\affiliation{\small{Department of Physics, Faculty of Science, Shahid Chamran University of Ahvaz, Ahvaz, Iran.}}

%%%%%%%%%%%%%%%%%%%%%%%%%%%%%%%%%%%%%%%%%%%%%%%%%%%%%%%%%%%%%%%%%%%%%%%%%

%%%%%%%%%%%%%%%%%%%%%%%%%%%%%%%%%%%%%%%%%%%%%%%%%%%%%%%%%%%%%%%%%%%

\begin{abstract}
Instead of solving Dirac (or Klein-Gordon) equation for a many body system, in this paper a variational method has been used to investigate the properties of two dimensional (2D) strongly interacting fermions and the results have been applied to 2D liquid $^3He$ as the only real fermion system.
Our results show that this variational method, known as lowest order constrained variational method, can be used to relativistic 2D fermion systems with a good accuracy.
In the case of 2D liquid $^3He$, Our calculations showed that at higher temperatures relativistic effects are more significant and quantum mechanical effects play a minor role. Also, we have found that in this system, as expected, relativistic effects are not considerable at low temperatures.

    %-------------------------------------------------------------------------------------------------------------
\end{abstract}

\pacs{}
\maketitle
{\bf Keywords}: Relativistic Effects, Dirac Equation, Klein-Gordon Equation, Strongly Correlated Systems, Two Dimensional (2D) Liquid $^3He$

%%%%%%%%%%%%%%%%%%%%%%%%%%%%%%%%%%%%%%%%%%%%%%%%%%%%%%%
\section{INTRODUCTION}
\label{Intro}
Investigation the properties of relativistic interacting fermion systems is one of the most important issues in various fields of science such as relativistic electron gas \text{\cite{LIu1}}, nuclear matter \text{\cite{Jiang2}}, chemical systems \text{\cite{Pyykko3}} and neutron stars \text{\cite{Hu4}}. Although relativistic effects are usually partial in known experimental systems and the Schr\"{o}dinger equation can be used to describe them with good accuracy, but these effects are significant in some systems such as electrons in heavy atoms (e.g. $Au$, $Cs$, where relativistic effects are considerable in molecule Bonds) \text{\cite{Engl5}}, light atoms (such as $Fe$, and $S$) \text{\cite{Michauk6}} and new-underlying 2D systems such as Graphene \text{\cite{Kim7,kivelson8}}.
%After the discovery of Graphene, theoretical and experimental studies indicate the existence of Dirac fermion in Graphene \cite{Kim7} and relativistic effects have been became very important in quantum matter confined in two spatial dimensions \cite{kivelson8}.
 %
In Graphene electrons can be considered as massless Dirac fermions obeying Dirac relativistic equation and their speed are two orders of magnitude smaller than the speed of light. Electrons in Graphene move in two spatial dimensions while relativistic quantum mechanics is configured for describing particles with three spatial degrees of freedom. Nowadays Dirac fermions have been also found in other systems such as the surface of topological insulators \text{\cite{DiVi9}}, quasi-2D systems \text{\cite{semenoff10}}, optical lattices \text{\cite{Hal11}} and Bismuth-based materials \text{\cite{NoVo12}}.
On the other hand, relativistic effects become more important in many-body systems theoretically, since one needs to solve Dirac relativistic equation in these systems. This problem, in fact, is a quantum electrodynamics (QED) one, but many attempts have been made to describe relativistic strongly correlated systems and different many-body methods including Green-function, Monte-Carlo, Coupled-Cluster, Landau Fermi-liquid model and etc. \text{\cite{Lin13}} have been presented to investigate these systems.
A relativistic strongly correlated system has its own difficulties in 2D as mentioned above and different works have been also done to describe these systems \text{\cite{Kim7, Lin13, Blas14}}.
These issues motivated us to investigate a 2D relativistic system in a variational manner. In this paper we have used a variational method known as lowest order constrained variational (LOCV) method \text{\cite{Bordbar15}} to investigate the properties of a 2D relativistic interacting fermion system. This method has been used in different correlated non-relativistic fermion systems with good accuracy, and the obtained results have a good agreement with experimental data \text{\cite{Bordbar16}}. Here, we have used this variational method to describe a relativistic many-body 2D problem considering the relativistic form of energy for particles instead of solving Dirac (and also Klein-Gordon) equation for this many body problem. The method has been applied in 2D liquid as the only real fermion many-body system and the results have been discussed and compared with theoretical and experimental data.
The structure of this paper is as follows. In next section, the applied variational method will be presented. Then the results of the method will be applied in 2D liquid $^3He$ and finally we have the Summery and Conclusion section.
%
%
%
%
%%%%%%%%%%%%%%%%%%%%%%%%%%%%%%%%%%%%%%%%%%%%%%%%%%%%%%%%%%%%%%%%%%%%%%%%%%%%%%%%%%%%%%%%%%%%%%%%%%%%%%%%%%%%%%%%%%%%%%%%%%
%
\section{Method: Lowest Order Constrained Variational Formalism}
We consider a two-dimensional system consists of $N$ interacting fermions whose distribution function at finite temperature is as follows,
\begin{equation} \label{eq1}
n({\bf k})=\frac{1}{e^{\beta \left(\varepsilon({\bf k}) -\mu \right)} +1} ,
\end{equation}
where $\beta=\frac{1}{K_BT}$ and $\mu$ is chemical potential obtained according to constrain of particle numbers, i.e.,
\begin{equation} \label{eq2}
N = \sum\limits_{\bf k} {n({\bf k})}  = \frac{{\nu A}}{{{{(2\pi )}^2}}}\int {{d}{\bf k}\,n({\bf k})},
\end{equation}
where $\nu$ is the single-particle level degeneracy ( $\nu=2$, for unpolarized fermions)and $A$ is the surface. Using Eq. \text{\ref{eq1}} we have,
\begin{widetext}
\begin{equation} \label{eq2a}
\rho  = \frac{\nu }{{2\pi }}\int\limits_0^\infty  {\frac{{kdk}}{{\exp \left[ {\beta \left( {{{({\hbar ^2}{c^2}{k^2} + {m^2}{c^4})}^{1/2}} - m{c^2} - \mu } \right)} \right] + 1}}},
\end{equation}
\end{widetext}
where $\rho$ is the number density of particles,
\begin{equation} \label{eq3}
\rho =\frac{N}{A}.
\end{equation}
In Eq. \text{ \ref{eq2a}} we have used the relativistic form for the single-particle energy as follows,
\begin{equation}\label{eq4}
\varepsilon (k) = {({\hbar ^2}{c^2}{k^2} + {m^2}{c^4})^{1/2}} - m{c^2}.
\end{equation}
In order to calculate thermodynamic properties of system, we use constrained variational method based on cluster expansion of energy functional. In this method the energy of system is written as follows,
\begin{equation}\label{Hamiltoni}
E =\frac{\langle\Psi|\Hat H|\Psi \rangle}{\langle\Psi|\Psi \rangle}=E_1+E_2+\dots
\end{equation}
where $\Hat H$  and $\Psi$ are the many-body Hamiltonian and wave function, respectively. We consider a many-particle trail wave function as
\begin{equation}\label{Fphi}
\Psi(1,2,...,N)=F(1,2,...,N)\Phi(1,2,...,N),
\end{equation}
where, $\Phi(1,2,...,N)$ is the Slater determinant of $N$ non-interacting
particles and $F(1,2,...,N)$ is the correlation operator containing the whole effects of interactions and has the cluster property, i.e., for each two subsets of $N$ interacting particles far from each other, $(i_1,\ldots,i_p)$ and $(i_{p+1},\ldots, i_N)$, we have
\begin{equation}\label{ClusterF}
  F(1,\ldots,N) \to F(1,\ldots ,p) F(p+1,\ldots,N)
\end{equation}
In Eq. \text{ \ref{Hamiltoni}}, $E_1$ and  $E_2$ are one- and two-body energies, respectively, calculated as follows:

One-body energy is the same as kinetic energy of non-interacting fermions written as follows \text{\cite{Bordbar15,Bordbar16}}:
\begin{equation}\label{eqE1}
{E_1} = \sum\limits_{\bf k } {n(\bf k)\varepsilon (\bf k)}.
\end{equation}
Here, $\varepsilon ({\bf k})$  is energy of a single particle whose relativistic form is as Eq. \text{ \ref{eq4}}. Therefore,
for relativistic many body fermion systems, at finite temperature, we have the one-body energy (per particle) as follows,
\begin{equation}\label{1body}
\frac{{{E_1}}}{N} = \frac{\upsilon }{{2\pi \rho }}\int\limits_0^\infty  {\frac{{(\sqrt {{\hbar ^2}{c^2}{k^2} + {m^2}{c^4}}  - m{c^2})}}{{{e^{\beta (\sqrt {{\hbar ^2}{c^2}{k^2} + {m^2}{c^4}}  - m{c^2} - \mu )}} + 1}}} kdk.
\end{equation}
It must be mentioned that for non-relativistic particles the single particle energy has following form and in Eqs \text{ \ref{eq4}}, \text{ \ref{eqE1}} and also other equations with summation (integration) of $n({\bf k})$, following equation must be substituted for single particle energy,
\begin{equation}\label{eqNRE1}
\varepsilon ({\bf{k}}) = \frac{{{\hbar ^2}{k^2}}}{{2m}}.
\end{equation}

Two-body energy, in our variational method, has the following form,
\begin{equation}\label{eqE2}
{E_2} = \frac{1}{{2N}}\sum\limits_{ij} {\left\langle {ij} \right|} W(12){\left| {ij} \right\rangle _a}
\end{equation}
where the subscript $"a"$ stands for anti-symmetric and  $\left| i \right\rangle$ is the one-particle wave function considered as a plane wave,
\begin{equation}\label{eqplanewave}
\left\langle {\bf r},\sigma| i \right\rangle  = \frac{1}{{\sqrt \Omega  }}{e^{i{{\bf{k}}_{\bf{i}}}.{\bf{r}}}}{\chi ^\sigma_i },
\end{equation}
such that $\chi ^\sigma_i $ the spinor of particle in state $i$ with spin $\sigma$. In Eq. \text{ \ref{eqE2}}, $W(12)(= W(r_{12}))$ represents the effective two-body potential,
\begin{equation}\label{Wgeneral}
\begin{array}{l}
W(12) = \frac{1}{2}{F^\dag }(12)\left[ {{\hat t}(1) + {\hat t}(2),F(12)} \right]\\
\,\,\,\,\,\,\,\,\,\,\,\,\,\,\,\,\,\,\,+ \frac{1}{2}\left[ {{F^\dag }(12),{\hat t}(1) + {\hat t}(2)} \right]F(12)\\
\,\,\,\,\,\,\,\,\,\,\,\,\,\,\,\,\,\,\, + {F^\dag }(12)V(12)F(12)
\end{array}
\end{equation}
where, the operator ${\hat t}(i)$ gives us the single particle energy,
\begin{equation}\label{singleenergyop}
\hat t(i)\left| i \right\rangle  = {\varepsilon _i}\left| i \right\rangle
\end{equation}
and ${\varepsilon _i}$ obtained from Eq. \text{ \ref{eq4}} ( Eq. \text{ \ref{eqNRE1}}) for relativistic (non-relativistic) particles.
After some relatively complicated algebra, for relativistic particles, we have,%
\begin{equation}\label{eqW}%
W(12)=W_1(12) + W_2(12) + W_3(12),%
\end{equation}
where,
\begin{widetext}
\begin{equation}\label{eqallw}
\begin{array}{l}
{W_{1}}(12) = \frac{{{\hbar ^2}}}{m}{(\nabla f(r))^2} + {f^2}(r)V(r)\\
{W_{2}}(12) = \frac{{{\hbar ^4}}}{{8{m^3}{c^2}}}\left\{ {8\nabla f(r).{\nabla ^3}f(r) + 6{\nabla ^2}f(r){\nabla ^2}f(r)} \right\}\\
{W_{3}}(12) = \frac{{{\hbar ^6}}}{{16{m^5}{c^4}}}\left\{ {30{\nabla ^2}f(r){\nabla ^4}f(r) + 12\nabla f(r).{\nabla ^5}f(r) + 20{\nabla ^3}f(r).{\nabla ^3}f(r)} \right\}.
\end{array}
\end{equation}
\end{widetext}
In the case of non-relativistic particles, we have only the first term \text{\cite{Bordbar15}} and two last terms are due to the relativistic form of single particle energy.
%
%where, the first term, $W_{NR}(12)$, is called  non-relativistic effective potential because its apparent form is like as the non-relativistic effective potential in LOCV method  \cite{Bordbar15}, and $W_{R1}(12)$ and $ W_{R2}(12)$ are first and second order relativistic corrections of effective potential, respectively. The details of Eq. \ref{eqW} are as,

In these equations $f(r)$ is two-body correlation function and $V(12)\equiv V(r)$ is the inter-particle potential energy which in the case of 2D liquid $^3He$ is the Lenard-Jones potential,
\begin{equation}\label{LJpot}
V(r) = 4\varepsilon \left\lfloor {{{(\frac{\sigma }{r})}^{12}} - {{(\frac{\sigma }{r})}^6}} \right\rfloor
\end{equation}
where, $\varepsilon {\rm{ =  10}}{\rm{.22\ K}}$ and $\sigma  = 2.556\,{\text{\AA}}$. In our calculations we have used the Jastrow approximation in which the correlation functions are defined as follows,
\begin{equation}\label{jastrow}
\begin{array}{l}
f(1.....N) = \prod\limits_{i < j} {f({r_{ij}}} )\\
f(1)=1\\
f(12) = f({r_{12}})\\
f(123) = f({r_{12}})f(r{}_{23})f({r_{31}})\\
\cdots
\end{array}
\end{equation}
After some algebra we have the following equation for two-body energy per particle
\begin{equation}\label{eqE2-2}
\frac{E_2}{N} = \frac{\rho }{2}\int\limits_0^\infty  {L(r)d{\bf r} \left[ {W(r)} \right]},	
\end{equation}
where,
%
%\begin{widetext}
%
\begin{equation}\label{eqL(r)}
L(r) =1-\frac{\nu}{4\pi^2\rho^2}(\gamma(r))^{\frac{1}{2}},
\end{equation}
%
%\end{widetext}
%
and
\begin{equation}\label{eqGAMA}
\gamma (r) = \int\limits_0^\infty  n(k){j_0} (kr)kdk.
\end{equation}
Here, $J_0(kr)$ is the spherical bessel function. Variation of Eq. \text{ \ref{eq2a}} with respect to correlation function, $f(r)$, leads to the following equation,
\begin{equation}\label{eqf(r)}
f''(r) + 2f'(r)\frac{{L'(r)}}{{L(r)}} - \frac{m}{{{\hbar ^2}}}(V(r) - 2\lambda )f(r)=0,
\end{equation}‎
where $\lambda$ is the Lagrange multiplier due to the normalization condition \text{\cite{Bordbar15,Bordbar16}}.
%
%\begin{equation}\label{eqNormalization}
%\xi  = \frac{1}{N}\sum\limits_{ij} {\left\langle {ij} \right|\left[ {L(r)} \right]}  - {f^2}(r){\left| {ij}. \right\rangle _a}
%\end{equation}‎
%
Two-body correlation function, $f(r)$, can be obtained by numerical methods and then we can calculate two-body energy using Eq. \text{ \ref{eqE2-2}}.
In order to obtain thermodynamic properties of system, we must calculate free energy,
\begin{equation}\label{eqFree}
F = E - TS‎.
\end{equation}‎
In this equation, $E$ is the total energy obtained from Eq. \text{ \ref{Hamiltoni}}, $T$ is the temperature and $S$ is the entropy obtained using the following equation \text{\cite{Bordbar15,Bordbar16}},
\begin{widetext}
\begin{equation}\label{eqEntropy}
%\begin{split}
S=-{k_B}\sum\limits_\varepsilon{(1-n(\varepsilon ))Ln(1-n(\varepsilon ))}+n(\varepsilon )Ln\,n(\varepsilon).
%\end{split}
‎\end{equation}‎
\end{widetext}
%
%%%%%%%%%%%%%%%%%%%%%%%%%%%%%%%%%%%%%%%%%%%%%%%%%%%%%%%%%%%%%%%%%%%%%%%%%%%%%%%
%%%%%%%%%%%%%%%%%%%%%%%%%%%%%%%%%%%%%%%%%%%%%%%%%%%%%%%%%%%%%%%%%%%%%%%%%%%%%
%%%%%%%%%%%%%%%%%%%%%%%%%%%%%%%%%%%%%%%%%%%%%%%%%%%%%%%%%%%%%%%%%%%%%%%%%%%%%
%
%
\section{RESULTS AND DISCUSSION} \label{NLmatchingFFtex}
In this section, the method mentioned in previous section is applied for evaluation the properties of two-dimensional liquid $^3He$. Firstly, some theoretical aspects of this system will be discussed.

\begin{figure}
  \centering
  \includegraphics[width=.5\textwidth]{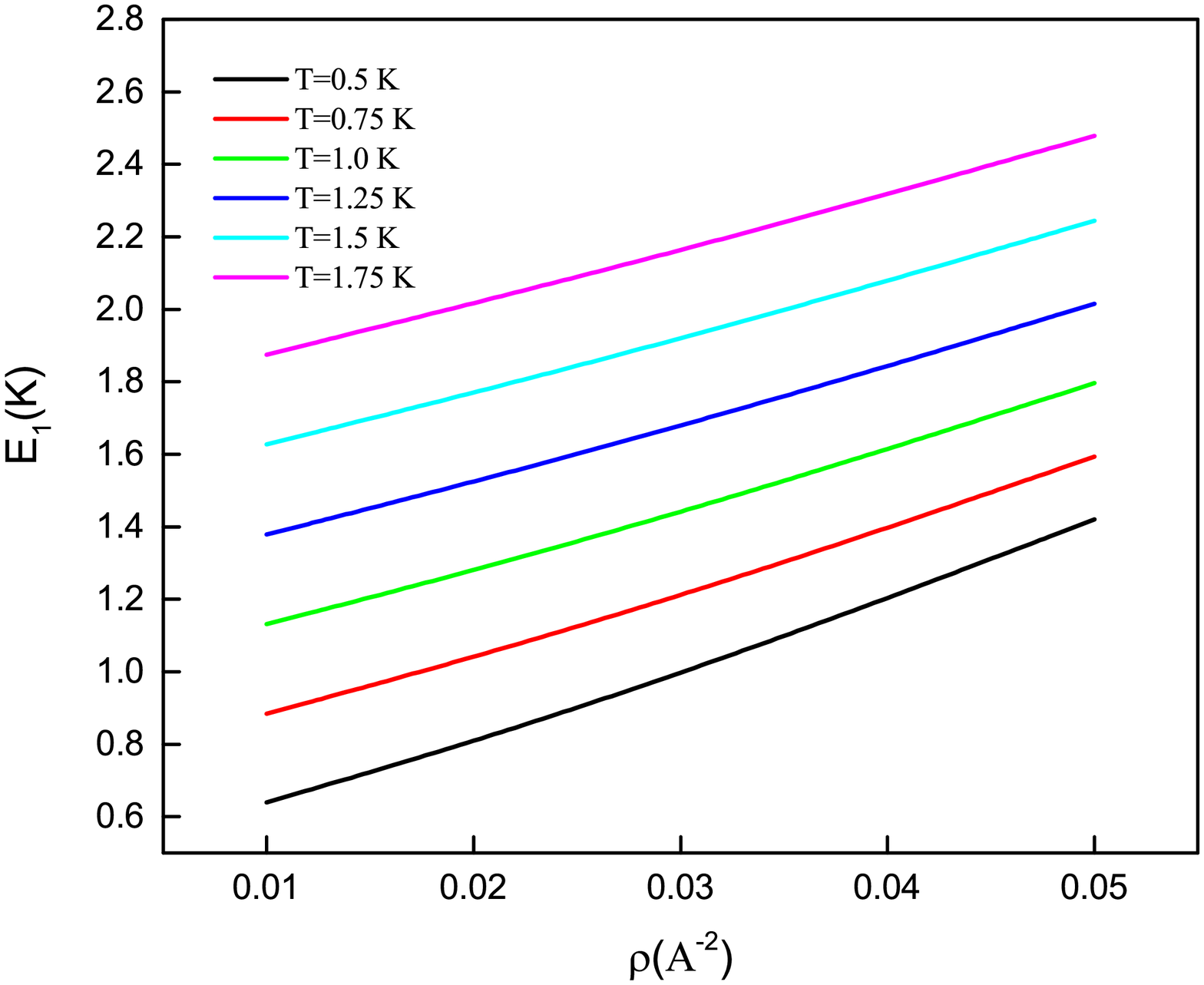}\includegraphics[width=.5\textwidth]{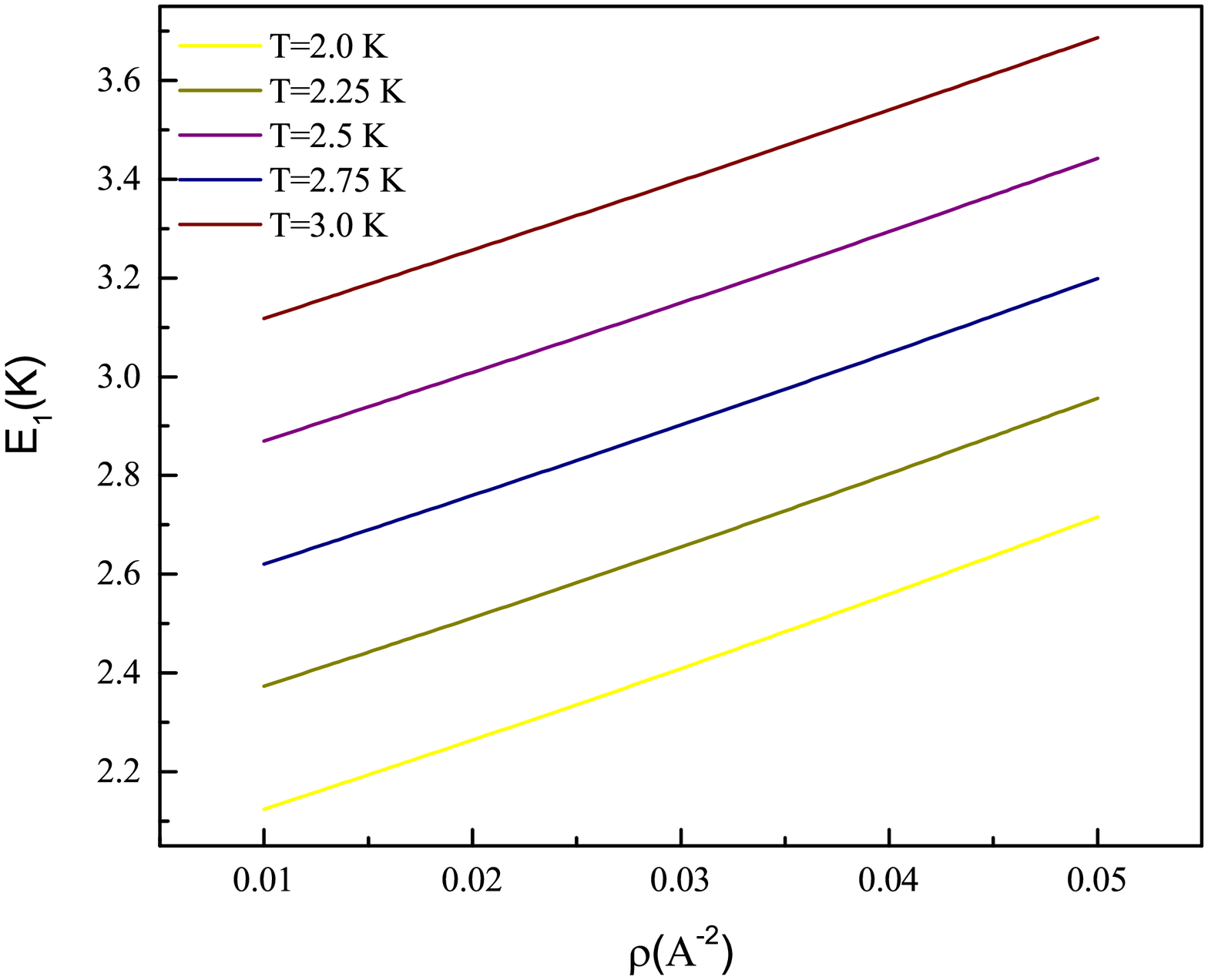}
  \caption{Kinetic (one-body) energy per particle of relativistic two-dimensional $^3He$ as a function of density for different temperatures.}\label{fig:E1ro}
\end{figure}

In Fig. \text{ \ref{fig:E1ro}}, the kinetic energy of relativistic two-dimensional (2D) liquid $^3He$ has been plotted as a function of density for different temperatures. As this figure shows, the kinetic energy per particle increases with increasing both density and temperature and this increment is almost monotonic.

\begin{figure}
  \centering
  \includegraphics[scale=.5]{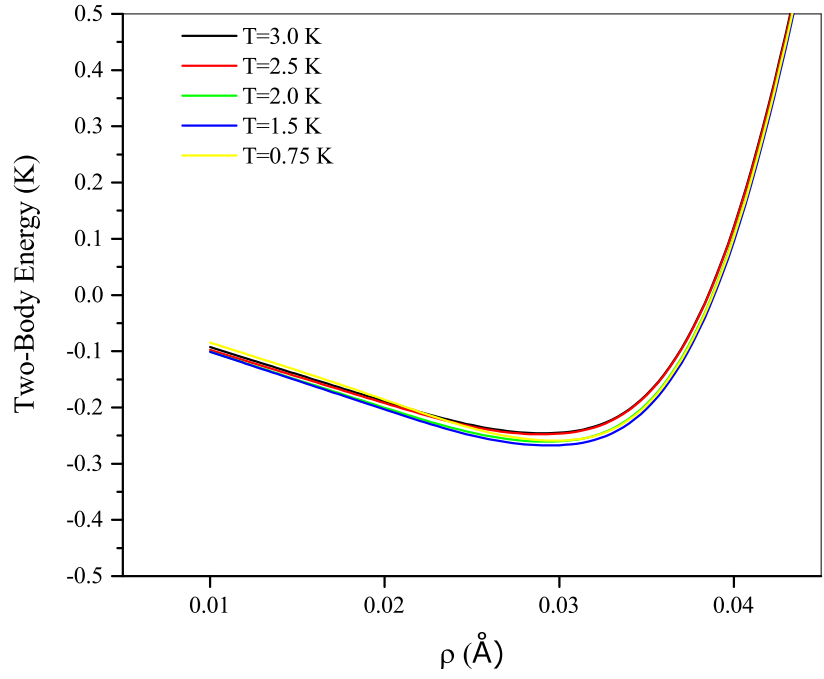}
  \caption{Two-body energy per particle of relativistic 2D liquid $^3He$ as a function of density at different temperatures.}\label{fig:E2ro}
\end{figure}

In Fig. \text{ \ref{fig:E2ro}}, two-body or interaction energy of 2D liquid $^3He$ has been plotted as a function of density for different temperatures. Our results show that in densities lower (greater) than $\rho  \approx {0.03{\text{\AA}}^{ - 2}}$, the two-body energy of 2D liquid $^3He$ decreases (increases) by density. In other words, by increasing density of particles, interaction energy becomes more effective because of decreasing the distance between particles.

\begin{figure}
  \centering
  \includegraphics[scale=.5]{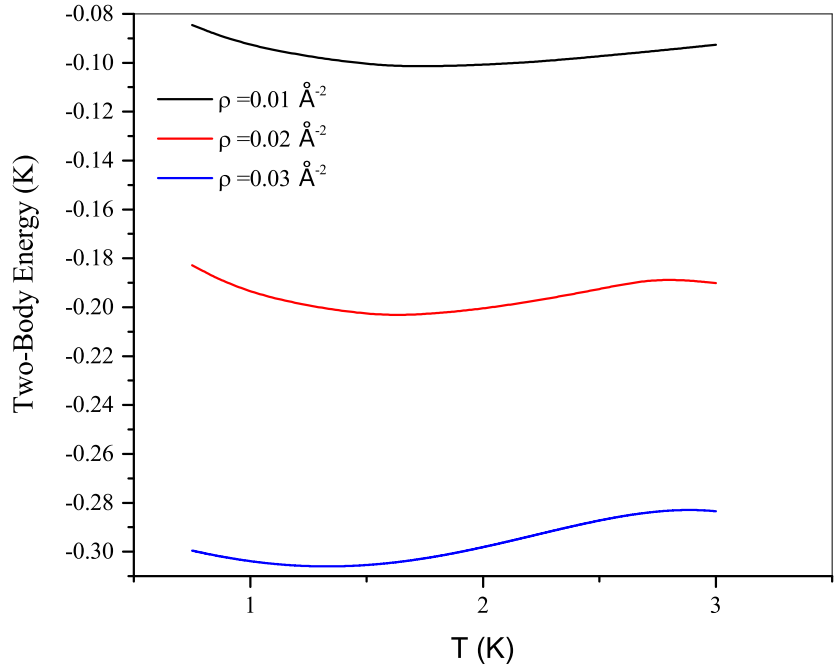}
  \caption{Two-body energy of relativistic 2D liquid $^3He$ as a function of temperature for different densities.}\label{fig:E2tem}
\end{figure}

Fig. \text{ \ref{fig:E2tem}} shows the interaction energy of 2D liquid $^3He$ as a function of temperature, considering the relativistic form of energy for particles. From this figure we can see that at temperatures below $T \approx 1.5\,K$ interaction energy of $^3He$ liquid decreases by increasing temperature, but in temperature greater than $T \approx 1.5\,K$, the two-body energy increases by increasing temperature. Our results show that the slope of this increment is approximately monotonic.

\begin{figure}
\centering
\includegraphics[width=.5\textwidth]{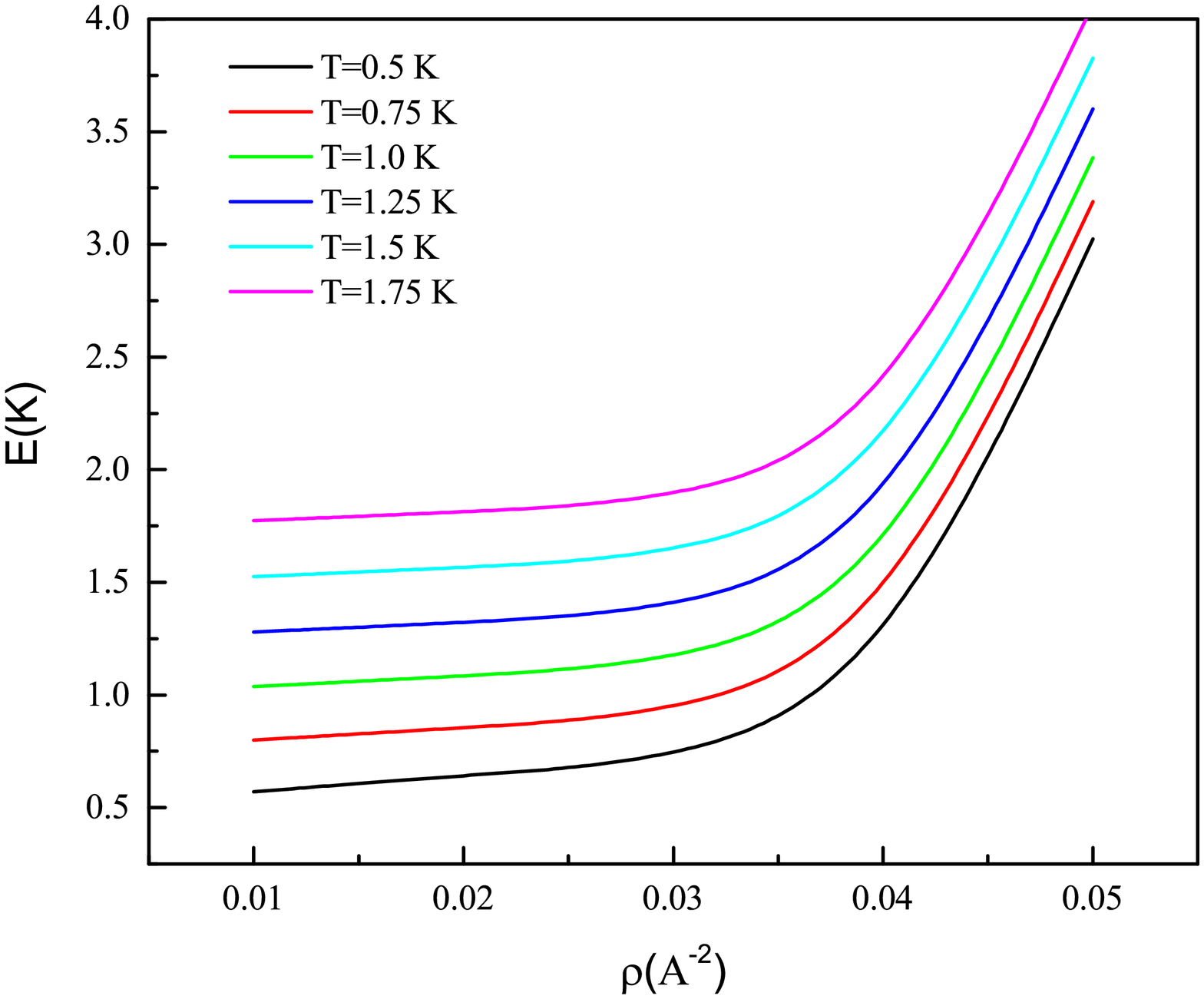}\includegraphics[width=.5\textwidth]{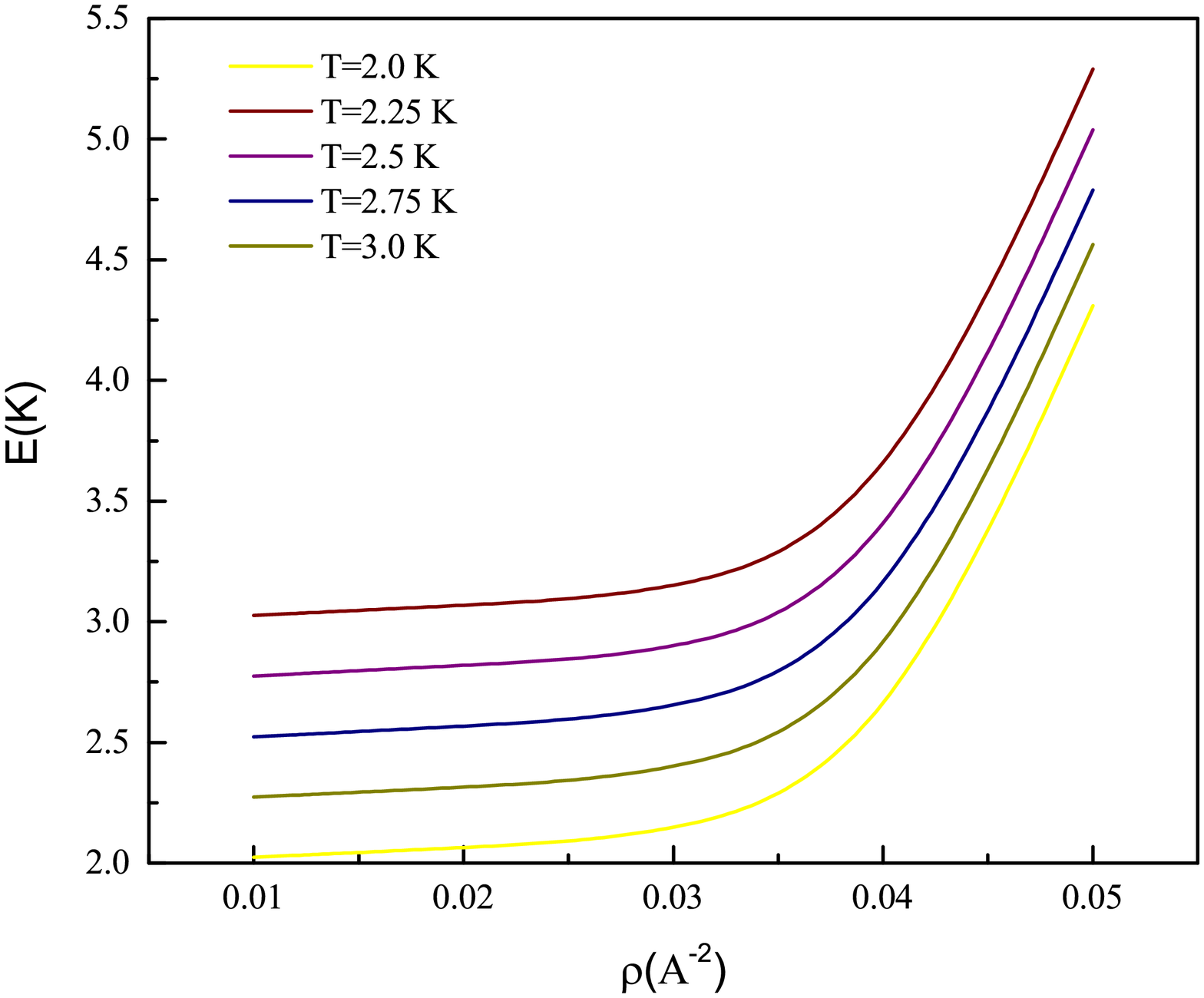}
\caption{Isothermals of total energy of relativistic 2D liquid $^3He$ as a function of density.}\label{fig:isothermal}
\end{figure}

Since the internal energy is the sum of kinetic and interaction energies, in Fig. \text{ \ref{fig:isothermal}}  the isothermal diagrams of total energy (per particle) of relativistic 2D liquid $^3He$ has been plotted as a function of density. As this figure shows, at a given temperature, the energy of relativistic 2D liquid $^3He$ increases by density. The increasing is more considerable at densities greater than $\rho \approx 0.03\ \text{\AA}^{-2}$. This means that two-body energy plays a significant role in the total energy of system.
As we have seen, two-body energy in our 2D relativistic variational model, includes three terms two of which are due to the considering relativistic form of single particle energy, Eq. \text{ \ref{eqW}} as well as the effect of single-particle energy effects (Eq. \text{ \ref{eq4}}) on the distribution function (Eq. \text{ \ref{eq2}}). On the other hand, it is believed that juxtaposing lower dimensionality for increasing interaction and quantum statistic changes some physical properties. Therefore, this behavior may be a results of all these reasons.

In Table \text{ \ref{tab:1}}, the ground state energy of relativistic 2D liquid $^3He$ has been presented for various densities. In order to have a better comparison, the results of some other works \text{\cite{NOvaco18, Miller19, Um20, Brami21}} are also presented in this table. As it can be seen, by increasing density, ground state energy of relativistic two-dimensional $^3He$ increases which this behavior has been observed at all temperatures. Our results show that there is no saturation point (minimum of energy as a function of density) for 2D liquid $^3He$ which is in agreement with all other theoretical works. It is be mentioned that the differences between our results and others may due to the influence of the substrate as well as three-body clusters not considered here.
Generally, the overall behavior of our results has a good agreement with the results of others. On the other hand, since total energy is the sum of kinetic (one-body) and interaction (two-body) energies (see Eq.\text{ \ref{Hamiltoni}}), results of internal energy show that the interaction energy plays a significant role in this behavior.
%which as mentioned before, is due to the relativistic corrections and also lowered dimensionality.

\begin{figure}
\centering
\includegraphics[width=.5\textwidth]{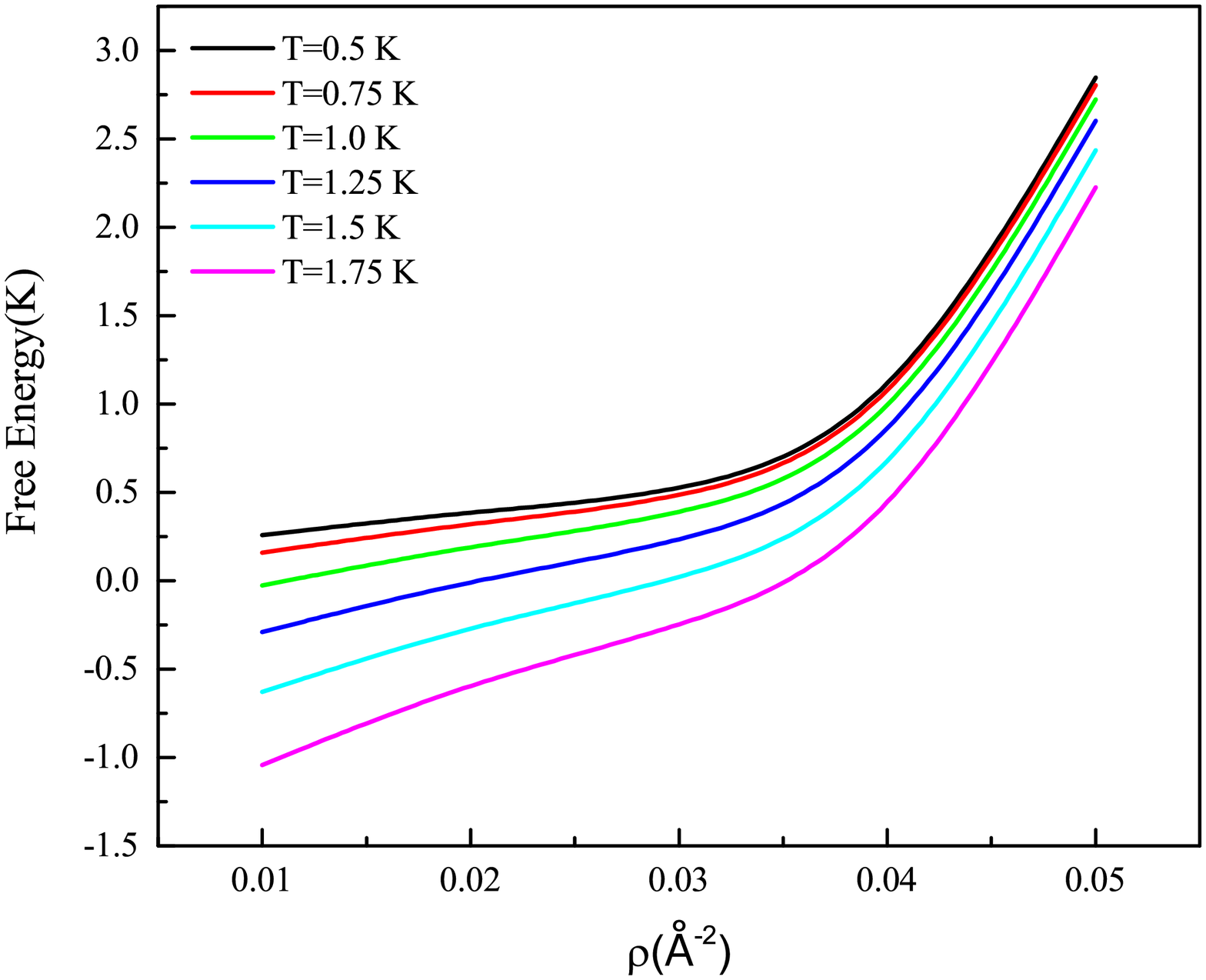}\includegraphics[width=.5\textwidth]{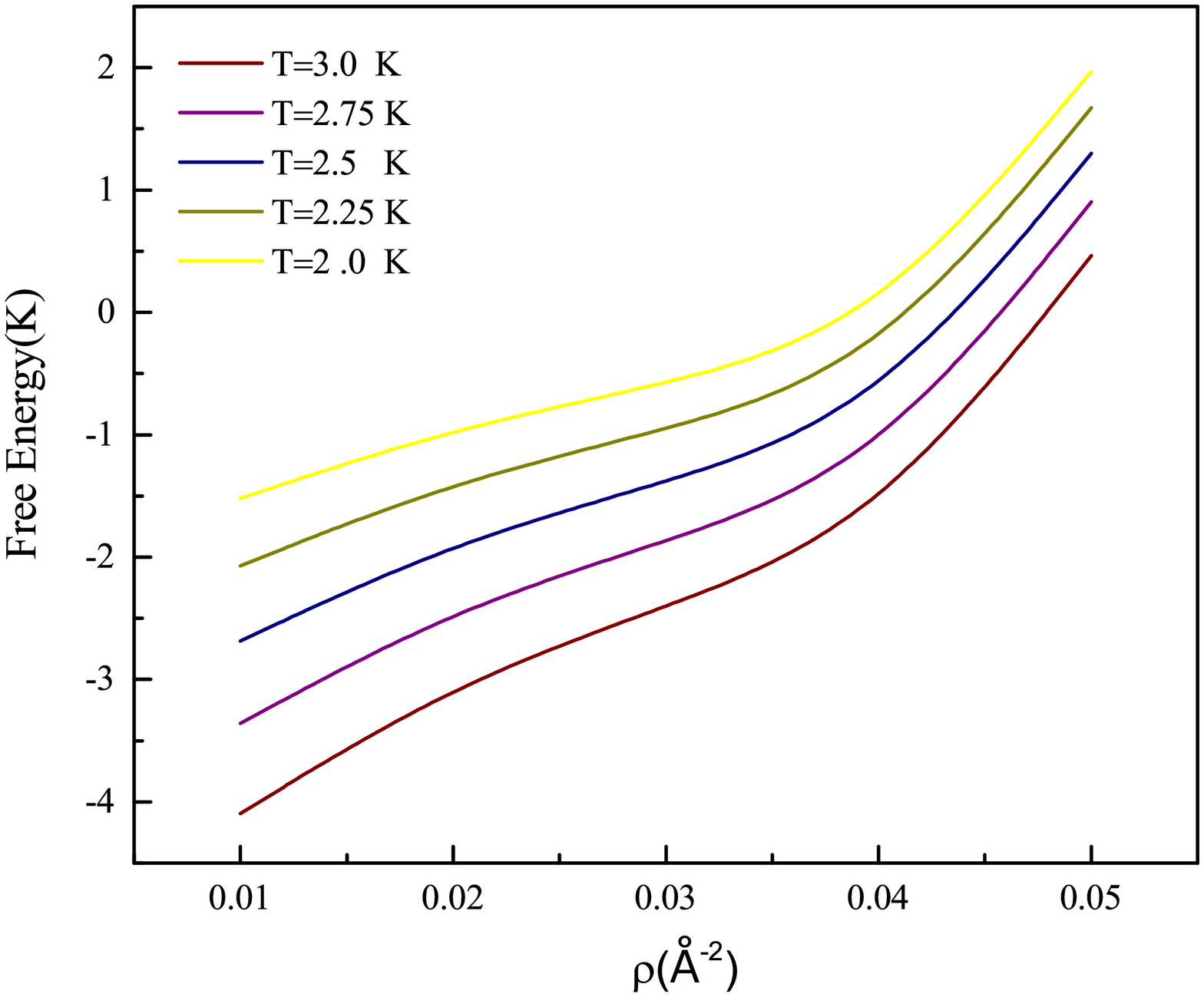}
\caption{Free energy of relativistic 2D liquid $^3He$ as a function of density at $T=2\ k$}\label{fig:FREE-RO}
\end{figure}

The results of free energy (as a function of density) have been plotted in Fig. \text{ \ref{fig:FREE-RO}}, in which the energy of particles has been considered in the relativistic form. Results have been calculated using $F = E - TS$ where $S$ and $E$ are entropy (Eq. \text{ \ref{eqFree}}) and internal energy of system (Eq. \text{ \ref{eq2}}), respectively. As it is clear from this figure, the free energy of relativistic 2D liquid $^3He$, at a given temperature, increases by increasing density. In this figure, again, there is no minimum point. It means that system has no bound state which is completely compatible with other theoretical \text{\cite{NOvaco18, Miller19}} and experimental works \text{\cite{Greywall22}}. Of course, in newer experimental works \text{\cite{sato23,sato24}} there are observed evidences of bound states in the two-dimensional $^3He$, but there is no such a state in any theoretical works up to now. Our results show that in all temperatures such an increasing behavior of the free energy can be observed. From this figure, we can also see that, the free energy of system decreases by increasing temperature. It means that in Eq. \text{ \ref{eqFree}}, $TS$ has a dominant part in free energy.

Also, in order to evaluate relativistic effects on this system, free energy of relativistic and non-relativistic at $T=2\ K$ has been presented in Table \text{ \ref{tab:2}}. Comparing these values shows that consideration relativistic corrections has a relative considerable effect on free energy of system. The results also showed that these effects increase with density and temperature.
%relativistic corrections of free energy in 2D liquid $^3He$ are of $10^{-1}$ order of magnitude and
%\textcolor{red}{
As we know the free energy is defined as Eq. \text{ \ref{eqFree}} and therefore these corrections are due to the total energy and entropy corrections which in turn is a result of the fact that effective potential is affected by this energy (see Eq.\text{ \ref{eqallw}})%
%}%
. Therefore, in order to investigate the effects of relativistic energy on systems, in Table \text{ \ref{tab:3}} the results of total energy of two-dimensional $^3He$ in relativistic state have been compared with that of non-relativistic case for different densities at $T=2\ K$.
In this table ${E_{NR}}$ and ${E_R}$ are the internal energy of non-relativistic and relativistic two-dimensional $^3He$ particles, respectively. As we have seen in Eq. \text{ \ref{eqW}}, in our variational method, considering the relativistic two dimensional strongly correlated systems (here, 2D liquid $^3He$) leads to three different parts in the effective potential. At a given temperature, the correction effects increase by increasing density.

In Table \text{ \ref{tab:4}}, the results of entropy of 2D liquid $^3He$  at $T=2\ K$, considering the relativistic form for single-particles energy, is compared with that of non-relativistic case for different densities. ${S_{NR}}$(${S_R}$) is the entropy of non-relativistic (relativistic) 2D liquid $^3He$. As we can see, for different densities the relativistic corrections are relatively considerable at high temperatures
Therefore, according to the last two tables and Eq. \text{ \ref{eqEntropy}}, relativistic effects in entropy and total energy lead to final effects on the free energy of systems.

\begin{figure}
\centering
\includegraphics[scale=.5]{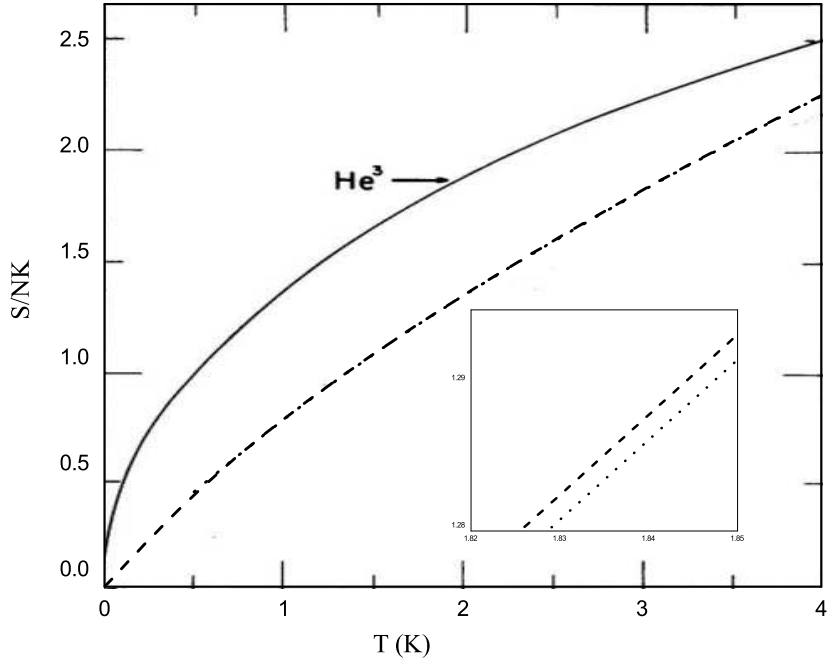}
\caption{Comparison the entropy of relativistic two-dimensional $^3He$ at $\rho = 0.028\,\text{\AA}^{-2} $ (dashed line) with experimental results (solid line) \text{\cite{Bretz25}}. The dotted line curve shows the results of entropy considering the non-relativistic form for single particle energy. The inner layout is drawn to note the differences between relativistic and non-relativistic entropies.
.}\label{fig:entropy}
\end{figure}

Results of entropy for relativistic 2D liquid $^3He$ has been plotted in Fig. \text{ \ref{fig:entropy}} at $\rho  = 0.028\,{{\text{\AA}}^{ - 2}}$. In order to better investigations, the experimental data \text{\cite{Bretz25}} have also been plotted. As we see from this figure, the overall behavior of our results have a good agreement with experimental data. It is be mentioned that the observed differences can be due to the effects of substrate which have not considered in this work. The results would be better by considering the three-body energy effects \text{\cite{Bordbar16}}.

\begin{figure}[h]
\centering
\includegraphics[scale=.6]{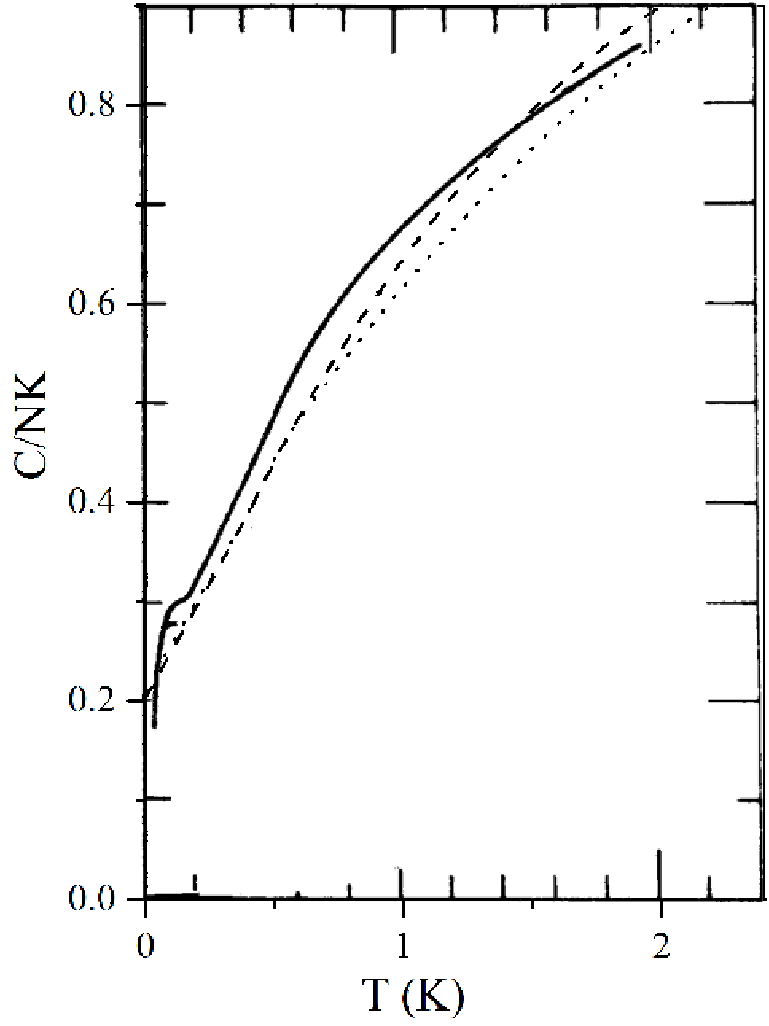}
\caption{Comparison the specific heat of relativistic two-dimensional liquid $^3He$ at $ \rho = 0.028\ \text{\AA}^{-2}$ (dashed line) with experimental results (solid line) \text{\cite{Bretz25}}. The dotted line shows the non-relativistic results for this density.}\label{fig:cv028}
\end{figure}

Another property is the specific heat of system. In Fig. \text{\ref{fig:cv028}}, the specific heat of 2D liquid $^3He$ has been plotted as a function of temperature for $\rho = 0.028\,{{\text{\AA}}^{ - 2}}$ and the experimental data \text{\cite{Bretz25}} have also been plotted for a better insight. As it is observed, the overall behavior of our results is in a good agreement with experimental data. From this figure one can also found that the relativistic results get closer to experimental data by increasing temperature, and relativistic corrections are more considerable at higher temperature. At high temperatures the kinetic energy is more important than the interaction energy and therefore, the relativistic effects have more considerable influence than quantum mechanical effects (which are mainly due to the interactions). Therefore, the results of relativistic corrections are more considerable by increasing temperature. At lower temperature the relativistic corrections are less considerable and mainly the quantum mechanical effects play more important roles.

In order to more evaluation, heat capacity of relativistic liquid $^3He$ in two densities at $T=4\ K$ are compared with experimental results. As we can see our results are in a good agreement with experimental results. Although in 2D liquid $^3He$ influence of relativistic corrections is small, but as it is clear from this table the values of specific heat in this relativistic 2D fermion system are in more agreement with experimental data \text{\cite{Bretz25}}.%

%----------------------------------------------------------------------------------
\section{Summary and Conclusions}%
In this paper a variational method has been used to investigate the properties of a 2D fermion system considering the relativistic form for the kinetic energy of single particles. The presented method has been applied to 2D liquid $^3He$ as the only real fermion system. Our results showed that the presented method can be used straightforward to relativistic two dimensional many body systems instead of solving Dirac or Klein-Gordon equation. The results of calculations applied in $^3He$ also showed that the relativistic corrections in this system are small, as expected. These corrections are more significant at higher temperature in which the kinetic energy of many body systems is more important than the interaction energy, and relativistic effects play a more important role rather than the quantum mechanical effects. The methods of this paper can be used to other relativistic 2D fermion many body systems including relativistic electron gas. %
%
%------------------------------------------------------------------------------
\acknowledgements{We wish to thank Ilam University
research council.}

%%%%%%%%%%%%%%%%%%%%%%%%%%%%%%%%%%%%%%%%%%%%%%%%%%%%%%%%%%%%%%%%%%%

%%%%%%%%%%%%%%%%%%%%%%%%%%%%%%%%%%%%%%%%%%%%%%%%%%%%%%%%%%%%%%%%%%%%%%%
\newpage
%\clearpage
%%%%%%%%%%%%%%%%%%%%%%%%%%%%%%%%%%%%%%%%%%%%%%%%%%%%%%%%%%%%%%%%%%%%%%%%%%%%%%%%%%
%
%
%-----------------------------------------------------------------------
%--------------------------------------------------------------------------
%=========================================================================
\newpage
\onecolumngrid
%==================
\clearpage
%%%%%%%%%%%%%%--------------------------------------
\clearpage
\centering
%%%=================================================================
\clearpage
%====================================
\clearpage
%%%============================
\clearpage
%=============================
\clearpage
%====================================
\clearpage
%====================================

%%%%%%%%%%%%%%%%%tables%%%%%%%%%%%%%%%%%%%%%%%%%%%%%%%%%%%%%%%%%%%%%
%===========================TABLES===============================
\begin{table}

\caption{Comparison the total energy of two-dimensional $^3He$ liquid in relativistic state with that of non-relativistic \text{\cite{NOvaco18, Miller19, Um20, Brami21}} case at $T=5\ K$.}%
\centering
\begin{center}
\begin{tabularx}{\textwidth}{|C|C|C|C|C|C|} \hline %{|c|c|c|c|c|c|c|} \hline
$\rho ({\text{\AA}}^{-2} )$ & \multicolumn{4}{|C|}{E${}_{NR}$(K)}  & $E_{R} (K)$\newline  \\ \cline{2-5}
 & \text{\cite{Brami21}} & \text{\cite{Um20}} & \text{\cite{Miller19}} & \text{\cite{NOvaco18}}  &   \\ \hline
0.005  & 0.175 & 0.075 & 0.10 & 0.22&  0.06117 \\ \hline
0.0075 & 0.25 & 0.12 & 0.14 & 0.35  &  0.1 \\ \hline
0.01   & 0.31 & 0.135 & 0.19 & 0.45 & 0.1443 \\ \hline
0.0125 & 0.45 & 0.15 & 0.23 & 0.58  & 0.1862   \\ \hline
0.015  & - & 0.175 & 0.28 & 0.68    & 0.23   \\ \hline
0.0175 & 0.525 & 0.21 & 0.34 & 0.84 & 0.28   \\ \hline
\end{tabularx}
\end{center}
\label{tab:1}
\end{table}
%===========================
\begin{table}
\caption{Comparison the free energy of 2D liquid $^3He$ in relativistic state with the free energy of non-relativistic case at $T=2\ K$.}
\begin{tabularx}{\textwidth}{CCC} %\hline
$\rho ({\text{\AA}}^{-2} )$ &$F_{R} (K)$ & $F_{NR} (K)$ \\ \hline
0.02 &-0.93668 & 1.10937 \\ %\hline
0.03 &-0.58832 & 1.25216 \\ %\hline
0.04 &-0.09854 & 1.60084 \\% \hline
0.05 &1.9743   & 3.56691 \\% \hline
\end{tabularx}
\label{tab:2}%
\end{table}
%----------------------------
\begin{table}
\centering
\caption{Comparison the total energy of 2D liquid $^3He$ in relativistic sate with that of the non-relativistic one.}\label{tab:3}
\begin{tabularx}{\textwidth}{CCC} %\hline
$\rho ({\text{\AA}}^{-2} )$ & $E_{R} (K)$ & $E_{NR} (K)\, \text{\cite{Bordbar17}} $  \\ \hline
0.02 & 2.0625 & 2.06476  \\ %\hline
0.03 & 2.1098 & 2.11156  \\ %\hline
0.04 & 2.3924 & 2.39435  \\ %\hline
0.05 & 4.3087 & 4.31054  \\ %\hline
\end{tabularx}
%\caption{comparison of total energy of 2D liquid $^3He$ in relativistic sate}\label{tab:3}
\end{table}
%================================
\begin{table}
\centering
\caption{Comparison of entropy of two-dimensional $^3He$ liquid in relativistic state with that of
 non-relativistic case $T=2\ K$.}
\begin{tabularx}{\textwidth}{CCC} %\hline
$\rho ({\text{\AA}}^{-2} )$ & S${}_{R}$ & S${}_{NR}$  \\ \hline
0.02 & 1.5027 & 1.50072  \\ %\hline
0.03 & 1.3517 & 1.34994  \\ %\hline
0.04 & 1.2481 & 1.24644  \\ %\hline
0.05 & 1.1696 & 1.16809  \\ %\hline
\end{tabularx}
\label{tab:4}
\end{table}
%===============================
\begin{table}
\centering
\caption{Comparison of heat capacity of two-dimension liquid $^3He$ in relativistic and non-relativistic state with experimental data \text{\cite{Bretz25}} at $T=4\ K$.}
\begin{tabularx}{\textwidth}{C C C C}

$\rho ({\text{\AA}}^{-2} )$  & $C_v\, (Non-Rel.)$ & $C_v\, (Rel.)$ & $C_v(Exp )$ \\ \hline
0.0415 & 1.018 & 1.015 & 0.862 \\ %\hline
0.0154 & 1.004 & 1.006 & 1.03  \\ %\hline
\end{tabularx}
\label{tab:5}
\end{table}
\end{document}